\documentstyle[12pt]{article}
\def\be{\begin{equation}}
\def\ee{\end{equation}}
\def\bea{\begin{eqnarray}}
\def\eea{\end{eqnarray}}
\newcommand{\CR}{\nonumber \\}
\newcommand{\A}{\alpha}
\newcommand{\B}{\beta}

\newcommand{\G}{\gamma}
\newcommand{\lm}{\lambda}
\newcommand{\LM}{\Lambda}

\newcommand{\pa}{\partial}

\newcommand{\cL}{{\cal L}}
\newcommand{\cF}{{\cal F}}
\newcommand{\ha}{\hat{a}}
\newcommand{\vt}{\vartheta}
\newcommand{\tl}{\tilde{l}}
\newcommand{\tm}{\tilde{m}}
\newcommand{\tn}{\tilde{n}}

\def\<{\langle}
\def\>{\rangle}
\def\non{\nonumber}
\def\lg{{\mbox{\cal g}}}

\begin{document}

\baselineskip=0.7cm

\renewcommand{\thefootnote}{\fnsymbol{footnote}}
\begin{titlepage}
\begin{flushright}
TU-508 \\
August 9, 1996
\end{flushright}

\bigskip

\begin{center}
{\Large \bf Exact and microscopic one-instanton calculations \\
in $N=2$ supersymmetric Yang-Mills theories}

\bigskip
\bigskip

Katsushi Ito\footnote{E-mail: ito@het.ph.tsukuba.ac.jp}

\medskip

{\it Institute of Physics, University of Tsukuba\\ Ibaraki 305, Japan}

\medskip

and

\medskip 

Naoki Sasakura\footnote{E-mail: sasakura@tuhep.phys.tohoku.ac.jp}

\medskip

{\it Department of Physics, Tohoku University \\ Sendai 980-77, Japan}

\end{center}

\bigskip

\bigskip

\begin{abstract}
We study the low-energy effective theory in $N=2$ super 
Yang-Mills theories by microscopic and exact approaches.
We calculate the one-instanton correction to the prepotential 
for any simple Lie group from the microscopic approach.
We also study the Picard-Fuchs equations 
and their solutions in the semi-classical regime for classical gauge groups
with rank $r\leq 3$. 
We find that for gauge groups $G=A_{r}$, $B_{r}$, $C_{r}$ ($r\leq 3$)
the microscopic results agree with those from the exact solutions.

\end{abstract}
\end{titlepage}

\renewcommand{\thefootnote}{\arabic{footnote}}
\setcounter{footnote}{0}

\section{Introduction}
The low-energy effective theory of $N=2$ supersymmetric gauge theory
in the Coulomb phase is determined by a single holomorphic
function, called the prepotential \cite{SEIZERO,GadWvP}.
Using holomorphy and duality, Seiberg and Witten \cite{sewi,sewi2}
have shown that the prepotential for the gauge group $G=SU(2)$ is 
completely determined by studying its singularities at strong coupling,
where massless monopole or dyon appears. 
The quantum moduli space of the low-energy effective theory is 
characterized by a family of elliptic curves, whose period becomes the
effective coupling of the theory. 
Examining these elliptic curves, one may calculate the non-perturbative
effects both in weak and strong coupling regions.
In particular, the exact solutions in the weak coupling regime leads to 
the non-perturbative instanton correction to the effective couplings,
while such a multi-instanton calculation is difficult and quite 
cumbersome in the microscopic theory.
{}From the viewpoint of the microscopic theory, 
the exact result provides a non-trivial and quantitative test
to the  method of instanton calculations.
On the other hand, 
the microscopic instanton calculation provides a non-trivial check
of the assumption of existence of massless monopole or dyon in the
strong coupling region, beyond the one-loop effect.

The microscopic instanton calculation in supersymmetric gauge 
theories \cite{ADSzero}-\cite{AMAKON} has some
good features. Firstly it is free from infrared 
divergences \cite{ROSVEN,AMAKON}. Therefore one 
may calculate its effect from the first principle.
Secondly the non-zero modes of the boson and fermion fields 
cancel with each other in the one-loop determinant \cite{ADDVEC}, 
and hence the problem reduces to the integration over finite number 
of zero-modes.
Finally, 
the holomorphic property of the supersymmetric gauge
theory \cite{AMAKON,SHIVAI,SEIONE} constrains 
the superpotential so severely that the lowest 
order calculation is enough to obtain a full order result.
In supersymmetric gauge theories, the above features will make it possible
to obtain correct instanton contributions by performing microscopic instanton 
calculations in the lowest order of gauge coupling constant $g \ll 1$, 
where such a semi-classical approximation would be reliable.
Thus one may compare the exact results with the microscopic instanton
calculations.

The microscopic instanton calculations have been investigated in the 
case of $N=2$ $SU(2)$ \cite{FINPOU,YUN}
and $SU(N_c)$ \cite{itsa} supersymmetric Yang-Mills theories (SYMs) 
at the one-instanton level, in the 
$SU(2)$ SYM \cite{DORKHOSYM,FUCGAB,DORKHOMUL} and 
supersymmetric QCD \cite{DORKHOQCD,AOYHAR,DORKHOQCDTWO} 
at the two-instanton level,
and in the $SU(2)$ SYM in the multi-instanton level \cite{DORKHOMUL}.
The microscopic instanton results have been consistent with the exact ones
so far except $SU(2)$ SQCDs with $N_f=3$ \cite{AOYHAR} and $N_f=4$
\cite{DORKHOQCDTWO},
where some discrepancies have been reported. 
But these discrepancies seem to come from the delicate differences 
between the observables of the exact and the microscopic instanton
calculations, and 
their results do not seem to contradict with the assumptions made by Seiberg
and Witten. 

In the previous paper \cite{itsa}, we performed the microscopic one-instanton 
calculation in $N=2$ $SU(N_c)$ SYM
and obtained the agreement between the exact and the microscopic 
one-instanton results for the $SU(3)$ SYM case.
In this paper we study the one-instanton correction to the prepotential
of $N=2$ SYM with arbitrary gauge groups from
both the microscopic and the exact calculations. 

The exact result for $G=SU(2)$ has been generalized to other gauge 
groups and SQCDs by introducing matter hypermultiplets
\cite{KlLeThYa}-\cite{ABOALI}.
For the computation of the prepotential, one needs to evaluate the 
contour integrals of a meromorphic differential over a 
hyperelliptic curve \cite{sewi,sewi2} in the weak coupling region.
An efficient approach to study this problem  is to use the Picard-Fuchs
differential equations and solve them near their singularities.
The Picard-Fuchs equations 
have been obtained for $SU(2)$ \cite{Cer} and 
$SU(3)$ SYM \cite{KLT} and $SU(2)$ SQCDs \cite{ItYa1,ItYa2,Oh}.
In the present work, we shall study the Picard-Fuchs equations for
classical gauge groups with rank $r\leq 4$ and examine their solutions
in the case of $G=BC_{2},A_{3}, B_{3}$ and $C_{3}$.

One of the technical difficulties in  the microscopic one-instanton 
calculation for general gauge groups is the group 
integration over the instanton configuration in the color space.
If the instanton calculation is done in the background of vanishing 
scalar vacuum expectation values \cite{ROSVEN,AMAKON},
the result of this integration turns out to be a numerical factor coming from 
gauge volume \cite{BER}. 
On the other hand, an $N=2$ SYM theory in the Coulomb phase has 
non-vanishing scalar vacuum expectation values, and the result of the 
group integration shows  non-trivial dependence on them. 
Although the group integration in the background of non-vanishing 
scalar vacuum expectation values was performed explicitly in 
$N=1$ $SU(N_c)$ SQCD with $N_c-1$ fundamental flavors \cite{COR,FUC},   
it looks quite formidable in general.  
In our previous work \cite{itsa}, 
we improve this technical difficulty by observing  that the 
result of the group integration should be a holomorphic function while
the integrand is a function of holomorphic and anti-holomorphic
variables.
We shall generalize this approach to the other gauge theories.

This paper is organized as follows;
In section 2, we will perform the microscopic one-instanton calculation
and will obtain the one-instanton correction to the prepotential in
the $N=2$ SYM theory with any simple Lie group.
In section 3, we will obtain the Picard-Fuchs equations for the 
exact solutions of $N=2$ SYM with some classical gauge groups and will 
calculate the one-instanton contributions to the prepotentials by 
solving them.
In section 4, we will discuss the relations between the scale parameters in 
the exact solutions and the dynamical scales in the microscopic
theories, and will compare the one-instanton corrections of the prepotential.
In Appendix, we will give the conventions on the Cartan-Weyl basis 
and the list of the Dynkin indices.    

\section{Microscopic one-instanton calculation} 

In this section, we will perform the microscopic one-instanton calculation
in the background of the non-vanishing scalar expectation values
in the lowest order of the gauge coupling constant. 
In this case, one needs to  use, for example,
the constraint instanton method \cite{AFF}
to perform an instanton calculation in a theory without a scale
invariance.
But, since the lowest order result will be enough to obtain a reliable 
full order result in a supersymmetric gauge theory
\cite{AMAKON,SHIVAI,SEIONE}, 
we do not need to worry about the higher order corrections 
coming from the constraint added to the action \cite{ADSzero,ADS}.
We will calculate a four-fermi correlation function of massless 
fermions in the microscopic theory \cite{SEIZERO,ADSzero,ADS} in the
one-instanton background. 
Then we will obtain the one-instanton correction to the prepotential.

\subsection{One-instanton configuration and integration measure in $N=2$
SYM theory}

We firstly discuss the zero-modes and the 
integration measure of the one-instanton configuration 
in the case of $N=2$ SYM theory with a general simple Lie group $G$.

The $N=2$ supersymmetric Yang-Mills theory with a simple gauge group $G$ 
contains an $N=1$ chiral
multiplet $\phi=(A,\psi)$ with the adjoint representation as well as
an $N=1$ vector multiplet $W_\alpha=(v_\mu,\lambda)$. 
The Lagrangian is\footnote{We follow
the conventions given in the text book by 
Wess and Bagger \cite{WESBAG}.}
\bea
{\cal L}&=&2 \int d^4 \theta \langle \phi^\dagger e^{-2gV} \phi e^{2gV}
\rangle_\lg + \frac{1}{2g^2}\left(\int d^2\theta \langle W^\alpha W_\alpha
\rangle_\lg + {\rm h.c.} \right), \nonumber \\
W_\alpha &=& -\frac{1}{8} \bar{D}^2 e^{-2gV} D_\alpha e^{2gV},
\label{orilag}
\eea
where $g$ is the gauge coupling constant, and $\<\cdot \>_\lg$ 
denotes the trace in the adjoint representation normalized as 
$\langle \cdot \rangle_\lg \equiv \frac{1}{k_D} {\rm Tr}_{adj}(\cdot)$
for convenience. Here $k_D$ denotes the Dynkin index of the adjoint 
representation of $\lg$, the  Lie algebra of $G$ (see Appendix). 
We will examine this Lagrangian in terms of component fields in the 
Wess-Zumino gauge. The euclidean Lagrangian of (\ref{orilag}) is given by  
\bea
{\cal L} &=& \frac{1}{2}\<  G_{\mu \nu} G_{\mu \nu} \>_\lg
+ 2 \< (D_\mu A)^\dagger D_\mu A \>_\lg + 
g^2 \< [A,A^\dagger ]^2 \>_\lg
+ 2i \< \Psi^\dagger M_F \Psi \>_\lg, \non \\
\Psi&=&\left({\lambda \atop \psi^\dagger}\right), 
\Psi^\dagger=(\lambda^\dagger, \psi), \non \\
M_F&=&\left( 
\begin{array}{cc}
-i\tau_\mu^- D_\mu & \sqrt2 g A \\
-\sqrt2 g A^\dagger & i\tau_\mu^+ D_\mu \\
\end{array}
\right),\non \\
\tau_\mu^\pm&=&(\sigma,\mp i), \CR
D_\mu&\equiv&\partial_\mu-ig v_\mu,
\label{comlag}
\eea
where $\sigma$ denotes the Pauli matrix.

Let us study the one-instanton solution of the gauge field.
In the lowest order of $g$, one can neglect the source terms from the
other fields, and the classical equation of motion of the gauge field 
$D_\mu G_{\mu\nu}=0$ has the instanton solution \cite{BPST,ADHM}. 
In the singular gauge, the instanton solution with unit topological
charge located at the origin is given by \cite{VAIZAK} 
\be
v_\mu=\frac2g \frac{\rho^2 \bar\eta_{a\mu\nu}x_\nu}{x^2(x^2+\rho^2)}
\Omega J^a\Omega^\dagger,
\label{inscon}
\ee
where  $\bar\eta_{a\mu\nu}$ and $\rho$ are the 't Hooft 
symbol \cite{GTH} and the instanton size, respectively.
The $\Omega$ is an element of the Lie group $G$, and the integration over it
is necessary to ensure the gauge invariance of the instanton calculation. 
The $J^a$ are the three generators of the $SU(2)$ obtained by 
embedding an $SU(2)$ into the Lie group $G$ which minimizes 
the classical value of the gauge kinetic term in the action. 
This is equivalent to taking the three generators
satisfying $[J^a,J^b]=i\varepsilon^{abc}J^c$ 
with a minimal value of $\<J^a J^a\>_\lg$.
Assuming $J^3$ is an element of the Cartan subalgebra $H$ of the Lie
algebra, it is obtained \cite{BERWEI} that 
\bea
J^3&=&\frac12 \sum_{i=1}^r \alpha^l_i H_i, \non \\
J^{\pm}&=&\frac1{\sqrt{2}}(J^1\pm iJ^2)=
\frac1{\sqrt{2}} E_{\pm \alpha^l}, 
\label{jis}
\eea
where $\alpha^l$ denotes one of the longest roots. More about our
conventions on the Cartan-Weyl basis are given in Appendix A. 
The norm of the 
generators takes the minimal value $\<J^a J^a\>_\lg=\frac12$, and 
the classical value of the gauge kinetic term is given by
\be
S_g=\frac{8\pi^2}{g^2}. 
\label{actgaukin}
\ee
Since $[J_3,E_\alpha] = \frac{1}{2} (\alpha^l,\alpha) E_{\alpha}$
and $(\alpha^l,\alpha)$ takes only the integral values $0,\pm 1,\pm 2$, 
all the generators in the Lie algebra $\lg$ are 
classified into the parts which have
the spin $0,\frac12,1$ representations of the embedded $SU(2)$ \cite{BERWEI}. 
The spin $1$ generators are those of the embedded $SU(2)$, and 
the number $d$ of the doublet pairs (spin $\frac12$) is related to the 
Dynkin index by 
\be
d=k_D-4,
\label{dtok}
\ee
which can be derived by evaluating $\<J^3 J^3\>_\lg$ in two 
ways \cite{BERWEI}.
We will write those doublet pairs more explicitly.
Define
\bea
E^{(\alpha^1)}&=&(E_{\alpha^1},E_{\alpha^1-\alpha^l}), \non \\
E^{(-\alpha^1)}&=& (E_{-\alpha^1+\alpha^l},-E_{-\alpha^1}) 
\label{douset}
\eea
for an $\alpha^1$ satisfying $(\alpha^l,\alpha^1)=1$.
With appropriate normalizations of $E_\alpha$'s ,
the $SU(2)$ generators are represented as
$[J_i,E^{(\pm \alpha^1)}]=E^{(\pm \alpha^1)} \frac12 \sigma_i$,
where $\sigma_i$ is the Pauli matrix.
We denote the set of $\alpha^1$'s labeling all the
two doublet pairs (\ref{douset}) by $\Delta^1(\alpha^l)$, which
has $d/2=(k_D-4)/2$ elements in total.

In the lowest order of $g$, the classical equation of motion of 
$\lambda$ is given by $\tau_\mu^- D^\mu \lambda=0$,
and the same for $\psi$.
The index theorem \cite{ATISIN} applied for this case 
tells that this equation has $k_D$ solutions.
These fermion zero-modes in the one-instanton background can be  
given explicitly by using the generators with spin $\frac12$ and $1$  of 
the embedded $SU(2)$ as follows (for $\Omega=1$):
\bea
{{\lambda^{SS}}_\alpha} &=& -\frac{\sqrt2 \rho^2}{\pi}
\frac{x_\mu x_\nu \eta_{c\nu\lambda}\bar\eta_{d\mu\lambda} (\sigma_c 
\xi^{SS})_\alpha}{x^2 (x^2+\rho^2)^2} J^d, \non \\
{\lambda^{SC}}_{\alpha} &=& \frac{\rho}{\pi} \frac{x_\nu 
\bar\eta_{c\nu\lambda}(\tau_\lambda^+  \xi^{SC})_\alpha}
{(x^2+\rho^2)^2}J^c, \non  \\
{{\lambda^{(\pm \alpha^1)}}_{\alpha}} &=& \frac{\rho i}{\sqrt2 \pi}
\frac{x_\mu({E^{(\pm \alpha^1)}} \tau_\mu^- \varepsilon )_\alpha
}{(x^2+\rho^2)^{3/2}\sqrt{x^2}} \xi^{(\pm \alpha^1)}, 
\label{fermizero}
\eea
where $\varepsilon$ denotes a $2$ by $2$ anti-symmetric tensor with 
$\varepsilon^{12}=1$, and 
we have used $\xi$ to label the gaugino zero-modes.
The $\lambda^{SS}$ and $\lambda^{SC}$  are called the supersymmetric
and the superconformal zero-modes, respectively, which can be obtained 
by applying the supersymmetry and the superconformal transformations to the 
gauge field (\ref{inscon}), respectively.
The label $\alpha^1$ runs through $\Delta^1(\alpha^l)$, and hence there
are $2+2+d=k_D$ zero-modes in (\ref{fermizero}). 

To define the integration measure, we will use the following norm:
\be
\parallel \phi(x) \parallel^2=\int d^4x 2 \< \phi(x)^\dagger \phi(x) \>_\lg,
\label{norm}
\ee
where the field $\phi(x)$ takes  values in the Lie algebra $\lg$. 
The integration measure on the group space is defined through the
following norm for an element $s$ in the Lie algebra $\lg$:
\be
\parallel s \parallel^2_{\mbox{\cal g}}=2 \< s^\dagger s  \>_\lg. 
\label{meaalg}
\ee

The bosonic zero-modes are the size $\rho$ and the location $x_0$
of the instanton as well as the location of the embedded $SU(2)$
in the Lie group $G$.
The change of the instanton configuration under the infinitesimal
shifts of those zero-modes were calculated explicitly, and 
the norms are given by \cite{BER}
\bea
\parallel A_\mu^{(\mbox{trans.})} \parallel  &=& \frac{2\sqrt2\pi}g, \non \\
\parallel A_\mu^{(\mbox{dilat.})} \parallel &=& \frac{4\pi}g, \non \\
\parallel A_\mu^{(\mbox{trip.})} \parallel &=& \frac{2\pi\rho}g, \non \\
\parallel A_\mu^{(\mbox{doub.})} \parallel &=& \frac{\sqrt2\pi\rho}g, 
\eea
where the trip. and doub. denote the infinitesimal shifts generated by
the triplet and doublet generators in $\lg$, respectively.
Since the number of the doublet zero-modes is related to the Dynkin
index by (\ref{dtok}), the bosonic integration measure is given by
\cite{BER,AMAKON,COR}
\be 
2^7\pi^{k_D}\mu^{2k_D}g^{-2k_D}
\int_0^\infty d\rho \rho^{2k_D-5}\int d^4x_0 \int_{G/G_s} d\Omega,
\label{measurenaive}
\ee
where we have introduced the Pauli-Villas regulator $\mu$, and 
$G_s$ is the stability group of the instanton
$G_s=\{ g \vert J^a= gJ^ag^\dagger , g\in G \}$.
Since we are interested only in a correlation function invariant under 
the space rotational transformation, this symmetry induces the invariance of 
the integrand under the color rotational symmetry 
generated by the embedded $SU(2)$ \cite{SHIVAINON}.
The integration over it results in the gauge volume 
$\mbox{Vol}(SU(2)_{adj}) =   2^3 \pi^2$,
where we have used the definition of the norm (\ref{meaalg}) 
in the  estimation of the gauge volume.

The fermionic zero-modes (\ref{fermizero}) are normalized to unity
under the norm (\ref{norm}). Hence the fermionic measure is given 
by \cite{GTH,NSVZ,AMAKON}
\be
\mu^{-k_D}\int d^2\xi^{SS}d^2\zeta^{SS}d^2\xi^{SC}d^2\zeta^{SC} 
\prod_{\alpha^1\in \Delta^1(\alpha^l)} d\xi^{(\alpha^1)}  d\zeta^{(\alpha^1)}
 d\xi^{(-\alpha^1)}  d\zeta^{(-\alpha^1)},
\label{fermimeasure}
\ee
where we have introduced $\zeta$ to label the $\psi$ zero-modes.

Gathering (\ref{actgaukin}),
(\ref{measurenaive}) and (\ref{fermimeasure}), 
the integration measure over the zero-modes of the instanton
configuration is given by 
\be
2^{10}\pi^{k_D+2}\Lambda_d^{k_D}g^{-2k_D} \int d^4x_0 \int_0^\infty 
d\rho \rho^{2k_D-5} \int_{G/SU(2)\times G_s} d\Omega
\int d^{2N_c}\xi  d^{2N_c}\zeta, 
\label{instmeasure}
\ee
where we have used $\int d^{2N_c}\xi  d^{2N_c}\zeta$
to denote the Berezin integration part in (\ref{fermimeasure})
for simplicity, and the dynamical scale is defined by
$\Lambda_d^{k_D}=\mu^{k_D}\exp\left( -\frac{8\pi^2}{g^2}\right)$.

\subsection{Instanton calculation with scalar vacuum expectation values}

Now we calculate the four fermion 
correlation function 
\be
\< {\psi_0^\dagger(x_1)}_{\dot\alpha}{\psi_0^\dagger(x_2)}^{\dot\alpha}
{\lambda_0^\dagger(x_3)}_{\dot\alpha'}{\lambda_0^\dagger(x_4)}^{\dot\alpha'}
\>
\label{foufer}
\ee
of the classically massless fermion fields in the 
microscopic theory \cite{SEIZERO,ADSzero,ADS}:
$
\psi_0^\dagger (x) \equiv \< \< A\> \psi^\dagger (x) \>_\lg$, 
$\lambda_0^\dagger (x) \equiv \< \< A\> \lambda^\dagger (x) \>_\lg$. 
We will need to evaluate the Yukawa terms to cancel the fermionic
zero-modes in the instanton measure (\ref{instmeasure}), and hence
the scalar vacuum expectation values will appear in the calculation. 

The potential term $\< [A,A^\dagger]^2 \>_\lg$ in (\ref{comlag})
has the classical flat directions
\bea
A_0&=&\sum_{i=1}^r a_i H_i, \non \\
A^\dagger_0&=&\sum_{i=1}^r \bar{a}_i H_i,
\eea
where $H_i$ are the generators in the Cartan subalgebra, and we take 
$a_i$ and $\bar{a}_i$ as independent variables to effectively use the 
holomorphy argument \cite{AMAKON,SHIVAI,SEIONE} in the instanton 
calculations \cite{itsa}.
For generic $a_i,\bar{a}_i$'s, 
the non-abelian gauge symmetry is completely broken to
the Cartan subalgebra $U(1)^r$ spanned by the $H_i$, and the system
is in the Coulomb phase.

As for the color space, the instanton resides in the $SU(2)$ subgroup 
minimally embedded to the gauge group $G$ \cite{BERWEI}
as was discussed in the previous subsection.
The integration over the embedding is needed to ensure the gauge
invariance of the instanton calculation. 
As can be shown by the global
gauge transformation, this is equivalent to the integration over the 
orientation of the scalar vacuum expectation values in the color 
space \cite{ADS,COR,FUC}.
Thus the group integration $\int d\Omega$  in (\ref{instmeasure}) can
be done by rotating the scalar expectation values, 
\bea
\<A\> &=& \Omega^\dagger A_0 \Omega, \non \\
\<A^\dagger \>&=& \Omega^\dagger A_0^\dagger \Omega,
\label{vacexp}
\eea  
while the embedding of $SU(2)$ itself is held fixed (The $\Omega$ in
(\ref{inscon}) is neglected.).

In the lowest order of $g$, the equation of motion of the scalar field
is given by $D_\mu^2 A(x)=0$.
The solution of this equation with the asymptotic value (\ref{vacexp})
is given by 
\be
A(x)=\frac{x^2}{x^2+\rho^2} \< A\>^t+\sqrt{\frac{x^2}{x^2+\rho^2}}\< A\>^d
+\< A\>^s,
\label{scasol}
\ee  
where $\< A\>^t,\< A\>^d,\< A\>^s$ denote the triplet, doublet and
singlet part of the $\< A\>$ concerning the embedded $SU(2)$.
Substituting this solution to the scalar kinetic term, we obtain
\bea
S_m&=& 8\pi^2 \rho^2 f(\<A\>,\<A^\dagger\>), \non \\
f(\<A\>,\<A^\dagger\>)&=&
\left\< \< A^\dagger\>^t\< A\>^t+\frac12 \< A^\dagger\>^d\<A \>^d 
\right\>_\lg.
\label{matact}
\eea

First we will consider the integration over the fermionic zero-modes.
In the lowest order of $g$, it turns out that the four supersymmetric 
zero-modes in the integration measure cancel with those appearing in the
massless fermion fields in (\ref{foufer}) \cite{SEIZERO,ADSzero,ADS}.
This is estimated by solving the equations of motion of those fields:
\bea
i\tau^{+\, \mu} D_\mu\psi^\dagger(x)
-\sqrt2 g [A^\dagger(x), \lambda^{SS}(x)]&=&0, 
\non \\
i\tau^{+\, \mu} D_\mu\lambda^\dagger(x)
+\sqrt2 g [A^\dagger(x), \psi^{SS}(x)]&=&0,
\eea
where we have 
substituted the supersymmetric fermionic zero-modes (\ref{fermizero})
into the source terms. These equations can be solved as
\bea
{\psi^\dagger(x)}^{\dot{\alpha}}&=&
\frac{g}{4\pi i} (\tau^{-\, \mu}\xi^{SS})^{\dot{\alpha}}D_\mu A^\dagger (x), 
\CR
{\lambda^\dagger(x)}^{\dot{\alpha}}&=&-
\frac{g}{4\pi i} (\tau^{-\, \mu}\zeta^{SS})^{\dot{\alpha}}D_\mu A^\dagger (x) 
\eea
by performing the supersymmetry transformation on the field 
$A^\dagger(x)$ \cite{ADSzero,ADS}. 
Since we are interested in the effective four-fermi vertex induced by 
an instanton, taking $x_i-x_0 \gg \rho$, we obtain
\be
 {\psi_0^\dagger(x_1)}_{\dot\alpha}{\psi_0^\dagger(x_2)}^{\dot\alpha}
{\lambda_0^\dagger(x_3)}_{\dot\alpha'}{\lambda_0^\dagger(x_4)}^{\dot\alpha'}
\sim (\pi g \rho^2 f)^4 \left( \frac12 \xi_{SS}^2\right)
\left(\frac12 \zeta_{SS}^2\right) S_F^4,
\ee
where we have denoted the product of the fermion propagators as
$
S_F^4 \equiv  
{{S_F}_{\dot\alpha}}^\beta(x_1-x_0){{S_F}^{\dot\alpha}}_\beta(x_2-x_0)
{{S_F}_{\dot\alpha'}}^{\beta'}(x_3-x_0){{S_F}^{\dot\alpha'}}_{\beta'}(x_4-x_0)
$
for simplicity.

The remaining fermion zero-modes appear in the Yukawa term 
$2\sqrt2 g i \< \psi(x)[A^\dagger(x),\lambda(x)]\>_\lg$.
Substituting the scalar field solution (\ref{scasol}) and the 
fermionic zero-modes (\ref{fermizero}) into it 
induces a bilinear term between  
$(\xi^{SC},\xi^{(\pm\alpha^1)})$ and $(\zeta^{SC},\zeta^{(\pm\alpha^1)})$
with a matrix $M(\<A^\dagger\>)$ depending
only on $\< A^\dagger \>$.

Hence, after the integration over the fermionic zero-modes and the instanton
size $\rho$, we obtain
\bea
\< \psi_0^\dagger\psi_0^\dagger\lambda_0^\dagger\lambda_0^\dagger \>
= -\frac{\Lambda_d^{k_D} \Gamma(k_D+2)}{g^{2k_D-4}2^{3k_D-3}\pi^{k_D-2}}
\int_{G/SU(2)\times G_s} d\Omega \frac{\mbox{det}M(\<A^\dagger\>)}
{f(\<A\>,\<A^\dagger\>)^{k_D-2}}
\int d^4x_0 S_F^4.
\label{midres}
\eea

\subsection{Group integration}

As will be discussed in the following subsection, the four fermi  
correlation function must be a holomorphic function of $A_0$ and 
must be independent of the anti-holomorphic variable $A_0^\dagger$.
Hence, taking the $A_0^\dagger \rightarrow \infty$ limit, the matter
action (\ref{matact}) will suppress the contributions other than the 
cases 
\be
\rho\sim0 \ \   \mbox{or} \ \  \< A\>^t,\< A\>^d\sim0.   
\label{intreg}
\ee
The former condition was previously discussed in the explicit
supersymmetric instanton calculation in the $N=1$ $SU(2)$ 
SQCD \cite{NSVZrho}. 
The fact that the contributions come only from the restricted region 
(\ref{intreg}) suggests that the group integration remaining in (\ref{midres})
is simple especially in supersymmetric gauge theories.
In this subsection, we will perform the group integration
\be
\int_{G/SU(2)\times G_s} d\Omega \frac{\mbox{det}M(\<A^\dagger\>)}
{f(\<A\>,\<A^\dagger\>)^{k_D-2}}
\ee
by generalizing the method 
we took in our previous paper for the $SU(N_c)$ case \cite{itsa}.
Since the result should be a holomorphic function of $A_0$,
we will obtain it by estimating its poles.

The poles will appear if $A_0$ is such that $f=0$ at some $\Omega$ . 
The condition $f=0$ will be understood as the holomorphic condition
$\< A\>^t=\< A\>^d=0$.
This is equivalent to 
\be
[\Omega J^a \Omega^\dagger, A_0]=0,
\label{conone}
\ee
where $J^a$ are the generators of the embedded $SU(2)$.
For general $\Omega$, the condition (\ref{conone}) will give two
conditions of $a_i$'s, and will not be related to the poles.
Thus we assume one of the ${J^a}'=\Omega' J^a {\Omega'}^\dagger$ is in the 
Cartan subalgebra $H$, say ${J^3}'\in H$, so that one of the conditions is
trivially satisfied.
Repeating the discussions in section 2.1, 
${J^i}'$ are given as in (\ref{jis}).
Thus the condition (\ref{conone}) becomes
\be
(\alpha^l,a)=0,   
\label{fincon}
\ee   
where $\alpha^l$ is one of the longest positive roots.
  
To estimate the order and the residue of the pole, we shift the
condition (\ref{fincon}) by an infinitesimally small parameter:
$(\alpha^l,a)=\epsilon$. Since $f=O(\epsilon)$ for $\Omega=\Omega'$, 
it turns out that only the following
small fluctuation of $\Omega$ is needed in the estimation of the pole:
\bea
\Omega &=& \exp(-i\sqrt{\epsilon}\delta g)\Omega', \CR
\delta g &=& \sum_{\alpha^1\in \Delta^1(\alpha^l)}
(x^{\alpha^1}E_{\alpha^1}+ {x^{\alpha^1}}^*
E_{-\alpha^1} + y^{\alpha^1}E_{\alpha^1-\alpha^l} + 
{y^{\alpha^1}}^*E_{-\alpha^1+\alpha^l}).
\eea
In fact, under this small fluctuation, $f$ remains $O(\epsilon)$: 
\be
f(\< A\> ,\< A^\dagger\> )=\frac{\epsilon}{2}(\bar{a},\alpha^l)+ 
\sum_{\alpha^1\in \Delta^1(\alpha^l)}
\epsilon (a,\alpha^1)
(\bar{a}_s,\alpha^1)\left( \vert x^{\alpha^1} \vert^2 + 
\vert y^{\alpha^1}\vert^2\right)+o(\epsilon).
\label{fes}
\ee
On the other hand, the leading term of det$M$ is $O(1)$:
\bea
(\mbox{Yukawa})&=& \sqrt2 gi\left( 
\frac12 (\alpha^l,\bar{a}) \xi_{SC}\varepsilon \sigma^3
\zeta_{SC} 
- \sum_{\alpha^1\in \Delta^1(\alpha^l)}
(\alpha^1,\bar{a}_s)(\xi^{\alpha^1}\zeta^{-\alpha^1}
+ \xi^{-\alpha^1}\zeta^{\alpha^1})\right)+o(1), \CR
\mbox{det}M&=&2^{k_D/2-3} g^{k_D-2} (\alpha^l,\bar{a})^2
\prod_{\alpha^1\in \Delta^1(\alpha^l)}(\alpha^1,\bar{a}_s)^2+o(1).
\label{det}
\eea
Here $\bar{a}_s$ denotes the singlet part of $\bar{a}$:
$\bar{a}_s \equiv \bar{a} - \frac12 \alpha^l(\bar{a},\alpha^l)$.

The integration measure of the group integration is defined through
the norm $\parallel \cdot \parallel_\lg$  in (\ref{meaalg}).
Since
$
\parallel \sqrt{\epsilon}\delta g \parallel^2_\lg 
=4\epsilon\sum_{\alpha^1\in \Delta^1(\alpha^l)}
(\vert x^{\alpha^1} \vert ^2 + \vert y^{\alpha^1} \vert ^2)$,
the integration measure is estimated as 
\be
d\Omega \sim 2^{2(k_D-4)} \epsilon^{k_D-4}  
\prod_{\alpha^1\in \Delta^1(\alpha^l)}
dx_1^{\alpha^1}dx_2^{\alpha^1}dy_1^{\alpha^1}dy_2^{\alpha^1} 
\label{epsmea}
\ee
in the lowest order of $\epsilon$, where 
$x^{\alpha^1}=x_1^{\alpha^1}+ix_2^{\alpha^1}$ and 
$y^{\alpha^1}=y_1^{\alpha^1}+iy_2^{\alpha^1}$.
Gathering (\ref{fes}), (\ref{det}) and (\ref{epsmea}) and integrating over 
$x$ and $y$, we obtain 
\be
\int_{G/SU(2)\times G_s} d\Omega \frac{\mbox{det} M}{f^{k_D-2}}
\sim
\frac{2^{\frac52 k_D-9}\pi^{k_D-4}g^{k_D-2}}{\Gamma(k_D-2)}
\times \frac1{\epsilon^2
\prod_{\alpha^1\in \Delta^1(\alpha^l)}(a,\alpha^1)^2}.
\label{resgro}
\ee

The full expression must have the similar poles for each $\alpha^l$
with $(a,\alpha^l)=0$. In fact, such a function can be constructed
for every simple Lie group.
Let us define a holomorphic function  $F_1^G(a)$ by 
\be
\int_{G/SU(2)\times G_s} d\Omega \frac{\mbox{det} M}{f^{k_D-2}}
=
\frac{2^{\frac52 k_D-9}\pi^{k_D-4}g^{k_D-2}}{\Gamma(k_D-2)}
F^G_1(a).
\label{finresgro}
\ee
An explicit form of the function $F_1^G(a)$ is given as follows:
\begin{itemize}
\item Simply laced Lie groups
 
All the root vectors have the same lengths. 
We have a general expression for these cases:
\be
F_1^G(a)=\frac{\sum_{\alpha\in \Delta_+(G)} 
\prod_{\alpha^0\in\Delta_+(G)\,:(\alpha,\alpha^0)=0}(a,\alpha^0)^2
\prod_{\alpha^1\in \Delta^1(\alpha)}(a,\alpha^1)(a,\alpha^1-\alpha)}
{\prod_{\alpha\in\Delta_+(G)} (a,\alpha)^2},
\label{fongen}
\ee
where $\alpha$ and $\alpha^0$ run over  the set of 
positive roots $\Delta_+(G)$.
The explicit expressions are also given below for the classical Lie groups
$A_r$ and $D_r$. 

\item $A_r$

The root vectors are described by using the orthogonal unit vectors $e_i\
$: $e_i-e_j$ $(i=1,\cdots,r+1\,;i\neq j)$.
Denoting $(a,e_i)$ as $a_i$ and substituting into (\ref{fongen}), 
we obtain
\bea
F_1^{A_r}(a)&=&
\frac{\sum_{i=1}^{r+1} 
\Delta^{r}(a_{1},\ldots, \widehat{a_{r}},\ldots, a_{r+1} )}
{2\Delta^{r+1}(a_{1},\ldots, a_{r+1})},\CR
\Delta^{m}(a_{1},\ldots, a_{m})&\equiv&\prod_{k<l}^m (a_{k}-a_{l})^2.
\label{fonar}
\eea

\item $B_r$ 

The root vectors are described by 
$\pm e_i$, $\pm e_i \pm e_j$ $(i,j=1,\cdots,r\,;i\neq j)$.
The latter ones are the longest roots. Denoting $(a,e_i)$ as
$a_i$, we obtain
\bea
F_1^{B_r}(a)&=&\frac{2\sum_{i=1}^r Q^{r-1}(a_{1},\ldots, 
\widehat{a_{i}},\ldots, a_{r})}{Q^r(a_{1},\ldots,a_{r})},\CR
Q^m(a_{1},\ldots, a_{m})&\equiv&\prod_{k<l}^m (a_{k}^2-a_{l}^2)^2.
\label{fonbr}
\eea

\item $C_r$

The root vectors are given by 
$\pm \sqrt2 e_i$, $\frac1{\sqrt2} ( \pm e_i \pm e_j)$ 
$(i,j=1,\cdots,r\,;\ i\neq j)$.
The former ones are the longest. Denoting $(a,e_i)$ as $a_i$, we obtain    
\be
F_1^{C_r}(a) = \frac{2^{r-2}}{\prod_{i=1}^r a_i^2}.
\label{foncr}
\ee
 
\item $D_r$

The root vectors are given by
$\pm e_i \pm e_j$ $(i,j=1,\cdots,r\,;i\neq j)$.
Denoting $(a,e_i)$ as $a_i$, we obtain
\be
F_1^{D_r}(a)=\frac{2\sum_{i=1}^r a_i^2 Q^{r-1}(a_{1},\ldots, 
\widehat{a_{i}},\ldots, a_{r}) }{Q^r(a_{1},\ldots,a_{r})}.
\label{fondr}
\ee
\item $G_2$ 

There are 6 positive roots:
$\alpha_1=(\sqrt{\frac23},0)$, 
$\alpha_2=(\frac1{\sqrt6},\frac1{\sqrt2})$,
$\alpha_3=(-\frac1{\sqrt6},\frac1{\sqrt2})$,
$\beta_1=(\sqrt{\frac32},\frac1{\sqrt2})$,
$\beta_2=(0,\sqrt2)$,
$\beta_3=(-\sqrt{\frac32},\frac1{\sqrt2})$. 
The $\beta_i$'s are the longest. We obtain
\be
F_1^{G_2}(a)=\frac9{\prod_{i=1}^3 (a,\beta_i)^2}.
\ee

\item $F_4$

There are 48 root vectors:
$\pm e_i$, $\pm e_i \pm e_j$ $(i,j=1,\cdots,4\,;i\neq j)$,
$\frac12 (\pm e_1\pm e_2\pm e_3 \pm e_4)$.
The second ones are the longest roots. Denoting $(a,e_i)$ as $a_i$, we obtain
\be
F_1^{F_4}(a)=\frac{2^5((a_1^2-a_2^2)^2(a_3^2-a_4^2)^2
+(a_1^2-a_3^2)^2(a_2^2-a_4^2)^2+(a_1^2-a_4^2)^2(a_2^2-a_3^2)^2)}
{\prod_{i<j}^4 (a_i^2-a_j^2)^2}.
\ee

\end{itemize}

\subsection{One-instanton correction to the prepotential}

Substituting (\ref{finresgro}) into (\ref{midres}),
we finally obtain the four-fermi correlation function as
\bea
\< \psi_0^\dagger\psi_0^\dagger\lambda_0^\dagger\lambda_0^\dagger \>
=
-\frac{ \Lambda_d^{k_D}}{g^{k_D-2}\pi^2 2^{k_D/2+6}}
\frac{\Gamma(k_D+2)}{\Gamma(k_D-2)} F^G_1(a) \int d^4x_0 S_F^4.
\label{finfou}
\eea

Now we will discuss the one-instanton correction to the prepotential.
Define the massless fields $\phi_i$ and $W^\alpha_i$ by  
\bea
\phi(x)&=&\frac1{g}\phi_i(x)H^i+(\mbox{massive fields}), \CR
W^{\alpha}(x)&=&W^{\alpha}_i(x)H^i+(\mbox{massive fields}).
\label{masfie}
\eea
In general,  $N=2$ low-energy effective action 
can be written of the form \cite{SEIZERO}: 
\be
{\cal L}_{eff}=\frac1{2\pi} {\rm Im}\left[ 
\int d^4\theta \left({\partial {\cal F}(\phi)\over \partial \phi_i} 
\bar{\phi}_i\right) + 
\int d^2\theta {1\over 2} \left({\partial^2{\cal F}(\phi)
\over \partial \phi_i
\partial \phi_j} W^\alpha_{\ i}W_{\alpha j} \right) \right].
\label{lowlag}
\ee
Substituting (\ref{masfie}) into (\ref{orilag}) results in
the classical part of the prepotential:
${\cal F}_{clas.}(\phi)=\frac{2\pi i}{g^2} \phi_i^2$.
In quantum case, including one-loop and non-perturbative instanton 
corrections, the prepotential takes the form
\be
{\cal F}(\phi)=\frac{\tau_0}2 \phi_i^2
+ \frac{i}{4\pi}\sum_{\alpha\in \Delta_+(G)} (\alpha,\phi)^2
\ln \frac{(\alpha,\phi)^2}{\Lambda_{d}^{2}}+
\sum_{n=1}^\infty {\cal F}^{inst.}_n(\phi) \Lambda_{d}^{k_Dn},
\label{quapre}
\ee
where the term ${\cal F}^{inst.}_n(\phi) \Lambda_{d}^{k_Dn}$ comes from 
the $n$-th instanton contribution.

Substituting $\phi_i\sim a_i+\sqrt2 \theta \psi_i$ and 
$W_i\sim -i \lambda_i$ into (\ref{lowlag}), 
one can see that the low-energy Lagrangian ${\cal L}_{eff}$ 
contains the four-fermi interaction
\be
\cL_{eff}=\cdots + 
\frac1{2^4\pi i} 
\frac{\partial^4 {\cal F}}
{\partial a_i \partial a_j \partial a_k \partial a_l}
\psi_i\psi_j\lambda_k\lambda_l+\cdots .
\ee
Since $\psi_0=a_i\psi_i$ and $\lambda_0=a_i\lambda_i$, the four-fermi
correlation function becomes
\be
\< \psi_0^\dagger\psi_0^\dagger\lambda_0^\dagger\lambda_0^\dagger \>
=
\frac{g^8}{2^6\pi i} a_i a_j a_k a_l 
\frac{\partial^4 {\cal F}(a)}{\partial a_i\partial a_j\partial a_k
\partial a_l} \int d^4x_0 S_F^4,
\ee
where we have performed the rotation to the euclidean space 
$(t\rightarrow -ix^4)$.
Comparing with the microscopic calculation (\ref{finfou}),
we obtain the one-instanton contribution to the prepotential as
\bea
\Lambda_d^{k_D}{\cal F}_1^{inst.}(\phi)=
-\frac{i \Lambda_d^{k_D}}{\pi 2^{k_D/2}} 
F^G_1(\phi).
\label{finone}
\eea
Here we have rescaled the fields as
$g a \rightarrow a$, $g^2 \psi_0^\dagger \rightarrow \psi_0^\dagger$,
$g^2 \lambda_0^\dagger \rightarrow \lambda_0^\dagger$,  
since the normalization of the fields are different between
(\ref{comlag}) and (\ref{lowlag}).

To compare this microscopic one-instanton result with the exact solutions, 
the all over normalization of the prepotential must be fixed. 
The classical part of the prepotential is not enough for
this purpose, because the bare coupling constant is not physical. 
On the other hand, the all over factor of the one-loop correction to 
the prepotential in (\ref{quapre})
does not depend on the bare coupling constant neither on the dynamical 
scale $\Lambda_d$.
In the sequel, we will always normalize the prepotential by this 
one-loop correction.

\section{Some exact results for classical gauge groups}

\subsection{The Picard-Fuchs equations for classical gauge groups}

In this section we study the quantum moduli space of the low-energy
effective theory of $N=2$ supersymmetric Yang-Mills theories
with classical gauge groups.
The low energy effective theory for a classical Lie group $G$
with rank $r$ is described by $N=2$ $U(1)^{r}$ vector multiplets
$\phi_{i}=(a_{i},\psi_{i})$, $W_{\A}^{i}=(\lambda_{i},v_{\mu}^{i})$ and the 
prepotential ${\cal F}(\phi)$, a holomorphic function of 
$\phi_{1}, \ldots, \phi_{r}$.
The quantum moduli space is then characterized by the family of 
hyperelliptic curves $C:y^{2}=f(x; \{ u_{i}\},\Lambda_{G})$, which is
parameterized by the gauge invariant Casimir invariants
$u_{i}={1\over i} <{\rm tr}\phi^{i}>$ and dynamically generated 
scale parameter $\Lambda_{G}$.
The fields $a_{i}$ and their dual fields
$a_{D i}={\partial {\cal F}\over \partial a_{i}}$ belong to the 
sections of $Sp(2r,{\bf Z})$ bundle over the quantum moduli space.
Their explicit form is obtained by the 
contour integral of the meromorphic differential $\lambda_{G}$ on the
hyperelliptic curve:
\bea
(\A_{i},a_{D})&=&\int_{B_{i}}\lm_{G}, \CR
(\lm_{i}, a)&=& \int_{A_{i}}\lm_{G},
\eea
where $(A_{i},B_{i})$ are certain homology cycles on $C$ with
the canonical intersection form $A_{i}\circ B_{j}=\delta_{ij}$,
$A_{i}\circ A_{j}=B_{i}\circ B_{j}=0$.
$\A_{i}$ are simple roots of $G$ and $\lm_{i}$ are fundamental weights of $G$.

The differential $\lambda_{G}$ is determined by the 
condition
\be
{\partial \lambda_{G} \over \partial s_{i}}=-\omega_{i}+{d f_{i}\over dx} dx,
\ee
where $s_{i}$ is certain polynomial of $u_{i}$'s and $\omega_{i}$ are
the basis of the holomorphic one-form on the curve.
The quantum moduli space contains singularities on which monopoles and/or
dyons become massless.
When the dynamical scale $\LM_{G}$ goes to zero, 
the effective theory becomes $N=2$ non-abelian
theory.
The prepotential receives the non-perturbative instanton corrections.

The hyperelliptic curve and the meromorphic differential are known 
for classical gauge groups and are given as follows: \\
For $A_{r}=SU(r+1)$ $(r\geq1)$ \cite{KlLeThYa}
\bea
y^{2}&=&P_{A_{r}}(x)^{2}-\LM_{A_{r}}^{2(r+1)}, \CR
\lm_{A_{r}}&=&{x {d P_{A_{r}}(x)\over d x}\over y} d x,
\label{hypar}
\eea
where
\be
P_{A_{r}}(x)=\prod_{i=1}^{r+1}(x-a_{i})
=x^{r+1}-\sum_{i=2}^{r+1} s_{i} x^{r+1-i}
, \quad \sum_{i=1}^{r+1}a_{i}=0.
\label{hypard}
\ee
For $B_{r}=SO(2r+1)$ ($r\geq2$) \cite{DaSu}
\bea
y^{2}&=&P_{B_{r}}(x)^{2}-\LM_{B_{r}}^{4r-2} x^{2}, \CR
\lm_{B_{r}}&=&{-x {d P_{B_{r}}(x)\over d x}+P_{B_{r}(x)}\over y} d x,
\eea
where
\be
P_{B_{r}}(x)=P_{r}(x)\equiv\prod_{i=1}^{r}(x^2-a_{i}^2)
=x^{2r}-\sum_{i=1}^{r} s_{2i} x^{2r-2i}.
\ee
For $C_{r}=Sp(2r)$ ($r\geq2$) \cite{ArSh}
\bea
x^{2} y^{2}&=&P_{C_{r}}(x)^{2}-\LM_{C_{r}}^{4r+4}, \CR
\lm_{C_{r}}&=&{- {d P_{C_{r}}(x)\over d x}\over y} d x,
\eea
where
\be
P_{C_{r}}(x)=x^{2}\prod_{i=1}^{r}(x^2-a_{i}^2)+\LM_{C_{r}}^{2r+2}
=x^{2}P_{r}(x)+\LM_{C_{r}}^{2r+2}.
\ee
For $D_{r}=SO(2r)$ ($r\geq 3$)  \cite{BrLa}
\bea
y^{2}&=&P_{D_{r}}(x)^{2}-\LM_{D_{r}}^{4r-4}x^{4}, \CR
\lm_{D_{r}}&=&{- {d P_{D_{r}}(x)\over d x}+2 P_{D_{r}}(x)\over y} d x,
\eea
where
\be
P_{D_{r}}(x)=P_{r}(x), \quad s_{2r}=t^{2}
\ee
and $s_{2},\ldots, s_{2r-2}, t$ are the gauge invariant order parameters.

In order to examine behavior of the periods $\Pi=(a^{i},a_{D}^{i})$ 
near the singularities, it is useful to 
study the differential equation (the Picard-Fuchs equation)
for the periods.
The solutions for the Picard-Fuchs equation have been investigated in the
case of $SU(N_{c})$ ($N_{c}=2,3$) \cite{KLT}, 
and  $SU(2)$ QCD with $N_{f}(\leq 3)$ massless \cite{ItYa1} and massive 
\cite{Oh} flavors.
In the following we will work on the Picard-Fuchs equations for classical
gauge groups.

We begin with the $A_{r}$ case, as discussed in ref. \cite{KLT}.
{}From the relation
\be
{\pa \lm_{A_{r}}\over \pa s_{i}}={x^{r+1-i} d x\over y}+{d\over dx}
\left( {x {\pa P_{A_{r}}(x)\over \pa s_{i}} \over y} \right) d x,
\label{eq:ar01}
\ee
we get
\be
{\pa^{2} \lm_{A_{r}}\over \pa s_{i}\pa s_{j}}=
-{x^{2r+2-i-j} d x\over y^{3}}+d (*).
\label{eq:ar02}
\ee
This implies the differential equations $\cL^{A_{r}}_{i,j;p,q}\Pi=0$, where
\be
\cL^{A_{r}}_{i,j;p,q}=\pa_{s_{i}}\pa_{s_{j}}-\pa_{s_{p}}\pa_{s_{q}}
\ee
for $i+j=p+q$.
Other types of differential equations may be obtained by considering
the total derivative
\be
{d \over d x}\left({x^{k}\over y}\right)
={k x^{k-1}\over y}-{x^{k} {d P_{A_{r}}\over d x} P_{A_{r}}\over y^{3}}.
\label{eq:ar1}
\ee
Since 
\be
x^{k} {d P_{A_{r}}\over d x}=(r+1) x^{r+1-k}
-\sum_{j=2}^{r+1} (r+1-j) s_{j} x^{r+1-j},
\ee
the r.h.s. of (\ref{eq:ar1}) may be expressed for $0\leq k\leq r-2$ 
in terms of derivatives
of $\lm_{A_{r}}$ up to total derivative, 
by using (\ref{eq:ar01}) and  (\ref{eq:ar02}).
Thus we may obtain a set of differential equations $\cL^{A_{r}}_{k}\Pi=0$, 
where
\bea
\cL^{A_{r}}_{0}&=& -(r+1)\pa_{s_{2}}\pa_{s_{r}}
+\sum_{j=2}^{r}(r+1-j)s_{j}\pa_{s_{r+1}}\pa_{s_{j+1}}, \CR
\cL^{A_{r}}_{k}&=& k\pa_{s_{r+2-k}}-(r+1)\pa_{s_{2}}\pa_{s_{r-k}}
+\sum_{j=2}^{r}(r+1-j) s_{j} \pa_{s_{r+2-k}}\pa_{s_{j}}.
\eea
For $k\geq r-1$, the r.h.s. of (\ref{eq:ar1}) does not give the second order
differential equation with respect to $s_{i}$. 
However, taking certain linear combination of 
$d({x^{r-1}\over y}),\cdots, d({x^{2r-1}\over y})$, one gets
the second order differential equation. 
Let us define this non-trivial equation as $\cL^{A_{r}}_{r-1}\Pi=0$.
$\cL^{A_{r}}_{r-1}$ is
determined modulo $\cL^{A_{r}}_{k}$ ($0\leq k\leq r-2$).
The differential operator $\cL^{A_{r}}_{r-1}$ contains 
the scale parameter $\LM_{A_{r}}^{2r+2}$
in contrast to other differential equations.
Although we may not obtain general expressions for $\cL^{A_{r}}_{r-1}$ 
in the present work,
we give examples for $A_{r}$ ($r\leq 4$) cases, which is given as follows:
\begin{itemize}
\item $A_{1}$
\be
\cL^{A_{1}}_{0}=-4(\LM_{A_{1}}^4-s_{2}^2)\pa_{s_{2}}^2+1.
\ee
\item $A_{2}$
\be
\cL^{A_{2}}_{1}=4 s_{2}^{2} \pa_{s_{2}}^{2}
-9(\LM_{A_{2}}^{6}
-s_{3}^{2})\pa_{s_{3}}^{2}+12 s_{2} s_{3}\pa_{s_{2}}\pa_{s_{3}}
+3 s_{3}\pa_{s_{3}}+1.
\ee
\item $A_{3}$
\bea
\cL^{A_{3}}_{2}&=&4(s_{2}^{2}+24 s_{4}) \pa_{s_{2}}^{2}
+9 s_{3}^{2} \pa_{s_{3}}^{2}
-16 (\LM_{A_{3}}^{8}-s_{4}^{2}) \pa_{s_{4}}^{2} \CR
& & +12 s_{2} s_{3}\pa_{s_{2}}\pa_{s_{3}}
-32 s_{2} s_{4} \pa_{s_{2}}\pa_{s_{4}}
+3 s_{3}\pa_{s_{3}}-16 s_{4} \pa_{s_{4}}+1.
\eea
\item $A_{4}$
\bea
\cL^{A_{4}}_{3}&=&4(s_{2}^{2}+20 s_{4}) \pa_{s_{2}}^{2}
+9 s_{3}^{2} \pa_{s_{3}}^{2}
-25 (\LM_{A_{4}}^{10} -s_{5}^{2}) \pa_{s_{5}}^{2} \CR
& & +12 s_{2} s_{3} \pa_{s_{2}}\pa_{s_{3}}
-32 s_{2} s_{4} \pa_{s_{2}}\pa_{s_{4}}
-8 s_{3} s_{4} \pa_{s_{3}}\pa_{s_{4}}
+20 s_{2} s_{5} \pa_{s_{2}}\pa_{s_{5}} \CR
& & +30 s_{3} s_{5} \pa_{s_{3}}\pa_{s_{5}}
+40 s_{4} s_{5} \pa_{s_{4}}\pa_{s_{5}} 
+3 s_{3}\pa_{s_{3}}-24 s_{4}\pa_{s_{4}}+15 s_{5} \pa_{s_{5}}
+1.
\eea
\end{itemize}
Notice that in the above results, the quantum correction 
appears only in the second derivative with respect to $s_{r+1}$.
For general $A_{r}$, one expects that the Picard-Fuchs
equation $\cL^{A_{r}}_{r-1}$ would take the form
\be
\cL^{A_{r}}_{r-1}=\cL^{A_{r}, classical}_{r-1}
-(r+1)^{2}\LM_{A_{r}}^{2r}\pa_{s_{r+1}}^{2},
\ee
where
$\cL^{A_{r}, classical}_{r-1}$ is certain differential operator which make 
the classical period $\oint \lm_{A_{r}}|_{\LM=0}=
\oint d x  x d \log (P_{A_{r}}(x))$ vanish.
The detailed analysis is left for future study.

Next we  consider the Picard-Fuchs equations for $B_{r}$-type gauge groups.
The meromorphic differential $\lm_{B_{r}}$ satisfies
\bea
\pa_{s_{2i}}\lm_{B_{r}}&=&-{x^{2r-2i}\over y} dx
-{d\over d x}\left( {x \pa_{s_{2i}} P_{r}(x)\over y}\right) dx, \CR
\pa_{s_{2i}}\pa_{s_{2j}}\lm_{B_{r}}
&=& -{x^{4r-i-j}P_{r}(x) \over y} dx+d(*). 
\label{eq:br0}
\eea
{}From (\ref{eq:br0}) and 
\be
d\left( {x^{2k+1}\over y}\right)
={2k x^{2k}\over y}-{x^{2k+1} {d P_{r}(x)\over d x}-x^{2k} P_{r}(x)\over y}
P_{r}(x),
\ee
we obtain a set of the Picard-Fuchs equations:
\bea
\cL^{B_{r}}_{i,j;p,q}\Pi&=&0, \quad i+j=p+q, \CR
\cL^{B_{r}}_{i}\Pi&=&0,  \quad i=0, \cdots, r-2,
\eea
where
\bea
\cL^{B_{r}}_{i,j;p,q}&=&\pa_{s_{2i}}\pa_{s_{2j}}-\pa_{s_{2p}}\pa_{s_{2q}}, \CR
\cL^{B_{r}}_{i}&=&-2 i \pa_{s_{2r-2i}}
+(2r-1)\pa_{s_{2}}\pa_{s_{2r-2-2i}}
-\sum_{k=1}^{r} (2r-2k-1) s_{2k} \pa_{s_{2k}}\pa_{s_{2r-2i}}.
\eea
For $B_{r}$ ($r\leq 4$), the differential equation $\cL^{B_{r}}_{r-1}\Pi=0$
which includes the scale parameter takes the form
\begin{itemize}
\item $B_{2}$
\be
\cL^{B_{2}}_{1}=4s_{2}^{2}\pa_{s_{2}}^{2}
        -(9\LM_{B_{2}}^{6}-16s_{2}s_{4})\pa_{s_{2}}\pa_{s_{4}}
        +16s_{4}^{2}\pa_{s_{4}}^{2}
        +8 s_{4}\pa_{s_{4}}+1.
\ee
\item $B_{3}$
\bea
\cL^{B_{3}}_{2}&=&4(s_{2}^{2}-105 s_{4})\pa_{s_{2}}^{2}
       +100 s_{4}^{2}\pa_{s_{4}}^{2}
       +36 s_{6}^{2}\pa_{s_{6}}^{2} \CR
     & &   +268 s_{2} s_{4}\pa_{s_{2}}\pa_{s_{4}}
       +24 s_{2} s_{6}\pa_{s_{2}}\pa_{s_{6}}
       -(25\LM_{B_{3}}^{10}+36 s_{4} s_{6})\pa_{s_{4}}\pa_{s_{6}} \CR
& &        +176 s_{4}\pa_{s_{4}}+24 s_{6}\pa_{s_{6}}+1.
\eea
\item $B_{4}$
\bea
\cL^{B_{4}}_{3}&=& 4 (s_{2}^{2}-448 s_{4}) \pa_{s_{2}}^{2}
+784 s_{2}^{2} \pa_{s_{2}}^{2}
+36 s_{6}^{2} \pa_{s_{6}}^{2}
+64s_{8}^2\pa_{s_{8}}^{2} 
 +1296 s_{2} s_{4} \pa_{s_{2}}\pa_{s_{4}} \CR
& & +24 s_{2} s_{6} \pa_{s_{2}}\pa_{s_{6}}
+32 s_{2} s_{8} \pa_{s_{2}}\pa_{s_{8}}
+304 s_{4} s_{6} \pa_{s_{4}}\pa_{s_{6}} 
-192 s_{4} s_{8} \pa_{s_{4}}\pa_{s_{8}}  \CR
& &  -(49 \LM_{B_{4}}^{14}-96s_{6} s_{8})\pa_{s_{6}}\pa_{s_{8}}
+1032 s_{4} \pa_{s_{4}}
+24 s_{6}\pa_{s_{6}}
+48s_{8}\pa_{s_{8}}
+1.
\eea
\end{itemize}

In a similar way we may derive the Picard-Fuchs equations for other classical
gauge groups.
Here we summarize the results.
For $C_{r}$ type gauge group, the Picard-Fuchs equations are
\bea
\cL^{C_{r}}_{i,j;p,q}\Pi&=&0, \quad i+j=p+q, \CR
\cL^{C_{r}}_{i}\Pi&=&0, \quad i=1, \ldots,r,
\eea
where
\bea
\cL^{C_{r}}_{i,j;p,q}&=&\pa_{s_{2i}}\pa_{s_{2j}}-\pa_{s_{2p}}\pa_{s_{2q}}, \CR
\cL^{C_{r}}_{i}&=&-(2i+1)\pa_{s_{2r+2-2i}}
+(2r+2) \pa_{s_{2}}\pa_{s_{2r-2i}}
-\sum_{k=1}~{r}(2r+2-2k) s_{2k}\pa_{s_{2k}}\pa_{s_{2r+2-2i}}, \CR
\eea
for $i=1,\ldots, r-1$.
$\cL^{C_{r}}_{r}$ for $C_{r}$ ($r=2,3, 4$) are 
\begin{itemize}
\item $C_{2}$
\be
\cL^{C_{2}}_{2}=4 s_{2}^{2}\pa_{s_{2}}^{2}
-(144\LM_{C_{2}}^{6}-16 s_{2} s_{4})\pa_{s_{2}}\pa_{s_{4}}
+(72 \LM_{C_{2}}^{6} s_{2}+16 s_{4}^{2}) \pa_{s_{4}}^{2}
+8 s_{4}\pa_{s_{4}}+1.
\ee
\item $C_{3}$
\bea
\cL^{C_{3}}_{3}&=&4 s_{2}^{2}\pa_{s_{2}}^{2}
+(16 s_{4}^{2}-384 \LM_{C_{3}}^{8}) \pa_{s_{4}}^{2}
+(36 s_{6}^{2}+128 \LM_{C_{3}}^{8} s_{4}) \pa_{s_{6}}^{2} \CR
& & +16 s_{2} s_{4}\pa_{s_{2}}\pa_{s_{4}}
    +24 s_{2} s_{6} \pa_{s_{2}}\pa_{s_{6}}
    +(256\LM_{C_{3}}^{8} s_{2}+48 s_{4} s_{6}) \pa_{s_{4}}\pa_{s_{6}} \CR
& & +8 s_{4}\pa_{s_{4}}+24 s_{6}\pa_{s_{6}}+1.
\eea
\item $C_{4}$
\bea
\cL^{C_{4}}_{4}&=&
4 s_{2}^{2} \pa_{s_{2}}^{2}
+16 s_{4}^{2} \pa_{s_{4}}^{2}
+36 s_{6}^{2} \pa_{s_{6}}^{2}
+(200 s_{6} \LM_{C_{4}}^{10}+64 s_{8}^2) \pa_{s_{8}}^{2}
+16 s_{2} s_{4} \pa_{s_{2}}\pa_{s_{4}} \CR
& & +24 s_{2} s_{6} \pa_{s_{2}}\pa_{s_{6}}
+(-800\LM_{C_{4}}^{10}+32 s_{2} s_{8}) \pa_{s_{2}}\pa_{s_{8}}
+48 s_{4} s_{6} \pa_{s_{4}}\pa_{s_{6}} \CR
& & 
+(600s_{2}\LM_{C_{4}}^{10}+64 s_{4} s_{8})  \pa_{s_{4}}\pa_{s_{8}}
+(400 s_{4}\LM_{C_{4}}^{10}+64 s_{6} s_{8}) \pa_{s_{6}}\pa_{s_{8}} \CR
& & +8 s_{4}\pa_{s_{4}}
+24 s_{6}\pa_{s_{6}}
+48 s_{8}\pa_{s_{8}}
+1 .
\eea
\end{itemize}

For $D_{r}$ ($r\geq3$), the Picard-Fuchs equations read
\bea
\cL^{D_{r}}_{i,j;p,q}\Pi&=&0, \quad i+j=p+q \CR
\cL^{D_{r}}_{i}\Pi&=&0, \quad i=0, \ldots,r-1
\eea
where
\bea
\cL^{D_{r}}_{i,j;p,q}&=&\pa_{s_{2i}}\pa_{s_{2j}}-\pa_{s_{2p}}\pa_{s_{2q}}, \CR
\cL^{D_{r}}_{i}&=&-(2i-1)\pa_{s_{2r-2i}}
+(2r-2) \pa_{s_{2}}\pa_{s_{2r-2-2i}}
-\sum_{k=1}~{r}(2r-2-2k) s_{2k}\pa_{s_{2k}}\pa_{s_{2r-2i}}, \CR
\eea
for $i=0,\ldots, r-2$.
The remaining differential operator $\cL^{D_{r}}_{r-1}$ 
for $D_{r}$ ($r=3,4$) take the form
\begin{itemize}
\item $D_{3}$
\bea
\cL^{D_{3}}_{2}&=&
(4 s_{2}^{2}-96 s_{4}) \pa_{s_{2}}^{2}+16 s_{4}^{2}\pa_{s_{4}}^{2}
        +(64 s_{2} s_{4}-72 s_{6}) \pa_{s_{2}}\pa_{s_{4}} \CR
      & &   +(-16 \LM_{D_{3}}^{8}+60 s_{2} s_{6}) \pa_{s_{2}}\pa_{s_{6}}
        +32 s_{4}\pa_{s_{4}}+6 s_{6} \pa_{s_{6}}+1.
\eea
\item $D_{4}$
\bea
\cL^{D_{4}}_{3}&=& 4(s_{2}^{2}-96 s_{4}) \pa_{s_{2}}^{2}
+144 s_{4}^{2} \pa_{s_{4}}^{2}
-36(\LM_{D_{4}}^{12}-s_{6}^{2})\pa_{s_{6}}^{2}
+64 s_{8}^{2} \pa_{s_{8}}^{2} 
 +272 s_{2} s_{4} \pa_{s_{2}}\pa_{s_{4}} \CR
& & +24 s_{2} s_{6}\pa_{s_{2}}\pa_{s_{6}}
+32 s_{2} s_{8}\pa_{s_{2}}\pa_{s_{8}}
+46 s_{4} s_{6}\pa_{s_{4}}\pa_{s_{6}}
-64 s_{4} s_{8} \pa_{s_{4}}\pa_{s_{8}}
+96 s_{6} s_{8}\pa_{s_{6}}\pa_{s_{8}} \CR
& & +200 s_{4}\pa_{s_{4}}
+24 s_{6} \pa_{s_{6}}
+48 s_{8} \pa_{s_{8}}+1.
\eea
\end{itemize}
So far we have written down the differential operators for classical 
gauge groups with rank $r\leq 4$.
We now check the consistency of the above Picard-Fuchs equations.
Since $B_{2}=C_{2}$ and $A_{3}=D_{3}$, the differential equations
should be equivalent though the curves look like different. 
For $B_{2}$ and $C_{2}$, one needs to change the basis of root system
\bea
a_{1}&=& {1\over\sqrt{2}}(\ha_{1}+\ha_{2}), \CR
a_{2}&=& {1\over\sqrt{2}}(\ha_{1}-\ha_{2}),
\eea
where $a_{i}$ and $\ha_{i}$ are the Higgs field in $C_{2}$ and $B_{2}$
curves, respectively.
This implies the change of variables
\bea
s_{2}&=&\hat{s}_{2}, \CR
s_{4}&=& -\hat{s}_{4}-{1\over4} \hat{s}_{2}^{2}.
\eea
One can check the Picard-Fuchs equations for $B_{2}$ and $C_{2}$ 
are equivalent if the scale parameters have a relation 
\be
\LM_{B_{2}}^{6}=-16\LM_{C_{2}}^{6}.
\label{btwandctw}
\ee
For $A_{3}$ and $D_{3}$, we find that both differential equations
are equivalent if the relations
\bea
s_{6}&=&-{1\over 8} \hat{s}_{3}^{2}, \CR
s_{4}&=& -\hat{s}_{4}-{1\over4} \hat{s}_{2}^{2}, \CR
s_{2}&=&-\hat{s}_{2}, \CR
\LM_{D_{3}}&=&\LM_{A_{3}}
\eea
hold. Here $s_{2i}$'s and $\hat{s}_{j}$'s are order parameters in $D_{3}$ and
$A_{3}$ curves respectively.

\subsection{Solutions of the Picard-Fuchs equations in the semi-classical
regime}
In this subsection we study the solutions of the Picard-Fuchs equations
in the semi-classical regime ($\LM\sim 0$) and 
compute the one-instanton contributions to the
prepotentials.
Firstly we consider the rank two case.
Since the $SU(3)$ Picard-Fuchs equations are studied in ref. \cite{KLT} using
Apell's generalized hypergeometric functions, we consider the $B_{2}$ case.
As in the case of $N=2$ $SU(3)$ massless SQCD \cite{ItYa2}, 
the $B_{2}$ Picard-Fuchs equations may not be written in the form of 
Apell's type. 
Let us rewrite the equations
by using Euler derivatives $\vt_{2i}=s_{2i}\pa_{s_{2i}}$:
\bea
\cL_{0}&=&3 s_{2}^{-2} (\vt_{2}^{2}-\vt_{2})
          +s_{4}^{-1} \vt_{4} (\vt_{4}-1-\vt_{2}), \CR
\cL_{1}&=& (2\vt_{2}+4\vt_{4}-1)^{2}-{9\LM_{B_{2}}^{6}\over s_{2}s_{4}}\vt_{2}\vt_{4}.
\label{eq:b2pf}
\eea
It is natural to introduce new variables
\be
x_{1}={s_{4}\over s_{2}^{2}}, \quad 
x_{2}=
{\LM_{B_{2}}^{6}\over s_{2} s_{4}}.
\ee
We construct the power series solution of (\ref{eq:b2pf}) 
around $(x_{1},x_{2})=(0,0)$. 
Let
\be
w(a,b;x_{1},x_{2})=\sum_{m,n\geq0} c_{m,n} x_{1}^{m+a} x_{2}^{n+b}
=s_{2}^{\A} s_{4}^{\B}+\cdots, 
\ee
be a formal power series of $x_{1}$ and $x_{2}$ with $c_{0,0}=1$.
Here  $\A=-2a-b$ and $\B=a-b$.
The indicial equations of (\ref{eq:b2pf}) read
\be
\beta(\beta-1-\alpha)=0, \quad (2\alpha+4\beta-1)^{2}=0.
\ee
These equations have two 
roots of multiplicity 2.
They are $(a,b)=(-1/6,-1/6)$, $(1/3,-1/6)$. 
{}From (\ref{eq:b2pf}),
$c_{m,n}$ obeys the recursion relations:
\bea
c_{m,n}&=& {-3 (m-1-n+a-b)(m-n+a-b-2)\over
            (m-n+a-b) (-(m+a)-1-2(n+b))} c_{m-1,n}, \CR
c_{m,n}&=& {-9 (m-n+a-b+1)(2(m+a)+(n-1+b))\over
             (6(n+b)+1)^2} c_{m,n-1}.
\eea
Thus we get two power series solutions
$w_{1}(s_{2},s_{4})=w(-1/6,-1/6,x_{1},x_{2})$, 
$w_{2}(s_{2},s_{4})=w(1/3,-1/6;x_{1},x_{2})$.
Two periods $a_{1}(s_{2},s_{4})$ and $a_{2}(s_{2},s_{4})$ are
expressed in terms of linear combinations of $w_{1}$ and $w_{2}$.
In the limit $\LM_{B_{2}}\rightarrow 0$, $a_{i}$ goes to $\phi_{i}$, where
$\phi_{i}$ is the solutions of 
\be
(x^{2}-a_{1}^{2})(x^{2}-a_{2}^{2})=x^{4}-s_{2} x^{2}-s_{4}=0.
\ee
This classical limit determines $a_{i}$. We find that
\be
a_{1}(s_{2},s_{4})= w_{1}(s_{2},s_{4}), \quad 
a_{2}(s_{2},s_{4})=i w_{2}(s_{2},s_{4}).
\label{eq:b21}
\ee
We may obtain the logarithmic type solutions $w_{D1}$, $w_{D2}$ 
by the Frobenius method:
\bea
w_{D1}(s_{1},s_{2})
&=&-3(\log s_{2}) w_{1}(s_{1},s_{2})-{s_{4}\over s_{2}^{3/2}}
+{3\over8} \LM_{B_{2}}^{6} {1\over s_{2}^{5/2}}
-{2\over3} \LM_{B_{2}}^{6} {s_{4}^{1/2}\over s_{2}^{7/2}}+\cdots, \CR
w_{D2}(s_{2},s_{4})&=& -\log(s_{2}s_{4}) w_{2}(s_{2},s_{4})
-{5\over3} {s_{4}^{3/2}\over s_{2}^{5/2}}
+{1\over 8} \LM_{B_{2}}^{6} {1\over s_{2}^{3/2} s_{4}^{1/2}}
+{1\over 2} \LM_{B_{2}}^{6} {s_{4}^{1/2}\over s_{2}^{7/2}}+\cdots .
\eea
The dual fields $a_{D}^{1}$ and $a_{D}^{2}$ are expressed as
$a_{D}^{1}=-w_{D1}+a_{11} w_{1}+a_{12} w_{2}$ and 
$a_{D}^{1}=-i w_{D2}+a_{21} w_{1}+a_{22} w_{2}$, where $a_{ij}$ are constants.
These constants are determined by evaluating the period integral.
For the computation of the instanton correction to the prepotential, 
however, explicit form of $a_{D}^{i}$ is not necessary.
Actually  the prepotential $\cF(a)$ satisfies the scaling equation \cite{Ma}
\be
8\pi i b_{1} s_{2}=\sum_{i=1}^{r} a_{i}{\pa\cF(a)\over a_{i}}-2\cF(a),
\label{eq:scal}
\ee
where $b_{1}$ is the coefficient of the 1-loop beta function
$b_{1}=k_{D}/16\pi^{2}$.
In the semi-classical regime, the exact 
prepotential ${\cal F}(a)$ has an expansion
\be
{\cal F}(a)=\frac{\tau'_{0}}{2} a^2
+ \frac{i}{4\pi}\sum_{\alpha\in \Delta_+(G)} (\alpha,a)^2
\ln \frac{(\alpha,a)^2}{\Lambda_{d}^{2}}+
\sum_{n=1}^\infty {\cal F}_n(a) \Lambda_{G}^{k_D n}.
\label{quaprecurve}
\ee
Therefore, to calculate the one-instanton correction term $\cF_{1}$, 
we need to know only explicit form of  $a_{i}$.
In the $B_{2}$ case, by inserting (\ref{eq:b21}) into (\ref{eq:scal}), 
we find that
\be
\LM_{B_{2}}^{6}\cF_{1}(a)=-{i\over4\pi}
{\LM_{B_{2}}^{6}\over 8} {2\over (a_{1}^2-a_{2}^2)^{2}}.
\label{exabtw}
\ee
In a similar way, one may perform exact calculation of the instanton 
calculation to the prepotentials for other gauge groups. 
In the present work, we have performed the explicit calculation 
up to rank 3 cases.
But as in the case of $A_{r}$, 
the matching condition for the scale parameter by making one of the 
vacuum expectation values of the Higgs fields large and the Weyl
group symmetry will determine $\cF_{1}$ explicitly, which will be 
discussed in the next section.

We next study the rank 3 case. 
Since $A_{3}$ and  $C_{3}$ cases may be treated in a  similar way, we discuss
the $B_{3}$ case in some detail.
Using the Euler derivatives, the differential equations read
\bea
\cL_{0}&=& -5 s_{2}^{-1}s_{4}^{-1}\vt_{2}\vt_{4}
           +s_{6}^{-1}\vt_{6}(3\vt_{2}+\vt_{4}-\vt_{6}+1), \CR
\cL_{1}&=& - s_{4}^{-1} \vt_{4}(3\vt_{2}+\vt_{4}-\vt_{6}+1)
           +5 s_{2}^{-2} \vt_{2}(\vt_{2}-1), \CR
\cL_{2}&=& 4\vt_{2}^{2}-4\vt_{2}+268\vt_{2}\vt_{4}+100\vt_{4}^{2}+76\vt_{4}
           +24\vt_{2}\vt_{6}-39\vt_{4}\vt_{6}+36\vt_{6}^{2}-12\vt_{6}+1 \CR
& & -420 {s_{4}\over s_{2}^{2}} \vt_{2}(\vt_{2}-1)
     -25{\LM_{B_{3}}^{10}\over s_{4}s_{6}}\vt_{4}\vt_{6}.
\eea
We introduce new variables $x_{1},x_{2},x_{3}$:
\be
x_{1}={s_{6}\over s_{2}s_{4}}, \quad
x_{2}={s_{4}\over s_{2}^{2}}, \quad
x_{3}={\LM_{B_{3}}^{10}\over s_{4} s_{6}}
\ee
and construct  power series solutions around $(x_{1},x_{2},x_{3})=(0,0,0)$
of the form
\be
w(a,b,c;x_{1},x_{2},x_{3})=
\sum_{l,m,n\geq0}d_{l,m,n} x_{1}^{l+a} x_{2}^{m+b} x_{3}^{n+c}
=s_{2}^{\A}s_{4}^{\B} s_{6}^{\G}+\cdots, 
\ee
where $\A=-a-2b$, $\B=-a+b-c$ and $\G=a-b$ and $d_{0,0,0}=1$.
The indicial equations become
\bea
& & \G(3\A+\B-\G+1)=0, \CR
& & \B(3\A+\B-\G+1)=0, \CR
& & 4\A^{2}-4\A+268\A\B+100\B^{2}+76\B
           +24\A\G-36\B\G+36\G^{2}-12\G+1=0,
\label{eq:indb31}
\eea
which have solutions
$(\A,\B,\G)=(1/2,0,0),(-1/2,1/2,0),(-1/4-\B/2,\B,1/4-\B/2)$.
$d_{l,m.n}$ obey the recursion relations:
\bea
d_{l,m,n}&=& {5\tl(l-1,m,n)\tm(l-1,m,n)\over
               \tn(l,m,n) (3\tl(l,m,n)+\tm(l,m,n)-\tn(l,m,n)+1)} d_{l-1,m.n},
\CR
d_{l,m,n}&=& {5 \tm(l,m-1,n) (\tm(l,m-1,n) -1)
\over          \tm(l,m,n) (3\tl(l,m,n)+\tm(l,m,n)-\tn(l,m,n)+1)}
d_{l,m-1,n}, \CR
d_{l,m,n}&=& {420 \tl(l,m-1,n) (\tl(l,m-1,n) -1)\over A(l,m.n)}
d_{l,m-1,n}+{25 \tm(l,m,n-1) \tn(l,m,n-1)\over A(l,m,n)} d_{l,m,n-1}, \CR
\eea
where $\tl(l,m,n)=-(l+a)-2(m+b)$, $\tm(l,m,n)=-(l+a)+(m+b)-(n+c)$,
$\tm(l,m,n)=(l+a)-(n+c)$ and
\bea
A(l,m,n)&=&4\tl(l,m,n)^{2}-4\tl(l,m,n)+268\tl(l,m,n)\tm(l,m,n)
 +
100\tm(l,m,n)^{2}\CR
& & +76\tm(l,m,n)
           +24\tl(l,m,n)\tn(l,m,n)-36\tm(l,m,n)\tn(l,m,n)\CR
& & +
36\tn(l,m,n)^{2}-12\tn(l,m,n)+1.
\eea

In order to determine $(a,b,c)$, we need to consider the classical limit
($\LM\rightarrow 0$).
In this limit $\pm a_{i}$ are the solution of 
$x^{6}-s_{2}x^{4}-s_{4}x^{2}-s_{6}=0$.
We thus find that the quantum $a_{i}$ becomes
\bea
a_{1}(s_{2},s_{4},s_{6})&=&
w({-1\over10},{-1\over5},{-1\over10};s_{2},s_{4},s_{6}),\CR
a_{2}(s_{2},s_{4},s_{6})&=&
{1+i\over \sqrt{2}}w({3\over20},{1\over20},{-1\over10};s_{2},s_{4},s_{6})
 +{\sqrt{2}(i-1)\over 8}
w({-7\over20},{11\over20},{-1\over10};s_{2},s_{4},s_{6})
, \CR
a_{3}(s_{2},s_{4},s_{6})&=&
{1-i\over \sqrt{2}}w({3\over20},{1\over20},{-1\over10};s_{2},s_{4},s_{6})
 +{\sqrt{2}(-i-1)\over 8}
w({-7\over20},{11\over20},{-1\over10};s_{2},s_{4},s_{6}). \CR
\eea
Using (\ref{eq:scal}), we find that one-instanton correction to the
prepotential is 
\bea
\cF_{1}=-{i\over 4\pi} {1\over 8} 
\left(
{1\over (a_{1}^{2}-a_{2}^{2})^{2} (a_{1}^{2}-a_{3}^{2})^{2}}
+{1\over (a_{1}^{2}-a_{2}^{2})^{2} (a_{2}^{2}-a_{3}^{2})^{2}}
+{1\over (a_{1}^{2}-a_{3}^{2})^{2} (a_{2}^{2}-a_{3}^{2})^{2}}\right) .
\label{exabth}
\eea
For $C_{r}$ type gauge groups, the results are as follows:
\be
\cF_{1}=\left\{
\begin{array}{cc}
\displaystyle{{i\over4\pi} {1\over a_{1}^{2} a_{2}^{2}}} 
& \mbox{for $C_{2}$,} \\
\displaystyle{{i\over4\pi} {1\over a_{1}^{2} a_{2}^{2}a_{3}^2}}
& \mbox{for $C_{3}$.}
\end{array}
\right.
\label{exac}
\ee
For $A_{3}$, we have the same result as in ref. \cite{itsa}, which 
is expected from the matching condition of the scale parameter.

\section{Matching conditions for scale parameters}

In this section, we will discuss the relations between the scale
parameters in the microscopic theories and those in the exact solutions 
for the classical gauge groups, 
and then will compare the one-instanton corrections to the 
prepotentials.
To do so, we will first discuss the matching conditions of the scale
parameters, and then will use the known relation 
in $N=2$ $SU(2)$ SYM theory \cite{FINPOU,itsa}.

When the gauge group is broken down from $G$ to $G'$ by the Higgs
mechanism, the matching condition of effective gauge couplings 
in the Pauli-Villas regularization scheme
leads to  the condition between the dynamical 
scales \cite{FINPOU}:
\be
\Lambda_{d,G}^{k_D}=\prod_i{m_i} \Lambda_{d,G'}^{k_D'},
\ee
where $m_i$'s denote the masses of the gauge bosons which become massive
by the Higgs scalar vacuum expectation values.  

Firstly we consider the  case $A_r \rightarrow A_{r-1}$. 
The root vectors of $A_r$ are given as in section 2.3. The reduction can 
be realized by considering a vacuum expectation value 
\be
a=b\sum_{i=1}^r e_i-rbe_{r+1}+\hat{a}, 
\quad (\hat{a},e_{r+1})=(\hat{a},\sum_{i=1}^r e_i)=0,
\ee
where $b \gg \hat{a}$. 
The massive gauge bosons are related to the roots 
$e_i-e_{r+1}\ \ (i=1,\cdots, r)$, and their mass is given by 
$\sqrt2 (a,e_i-e_{r+1})\sim \sqrt2 (1+r)b$.
Hence the matching condition is given by
\be
\Lambda_{d,A_r}^{2(r+1)}=2(1+r)^2b^2 \Lambda_{d,A_{r-1}}^{2r}.
\label{dmatar}
\ee
One can see easily that the one-instanton correction (\ref{fonar}), 
(\ref{finone}) is consistent with the matching condition (\ref{dmatar}).
On the other hand, the matching condition in the exact solution can be 
derived from the hyperelliptic curve most easily.
Substituting 
\be
a_i=(a,e_i)=b+\hat{a}_i\ (i=1,\cdots,r),\ \ a_{r+1}=(a,e_{r+1})=-rb
\ee
into (\ref{hypar}), (\ref{hypard}), one obtains the matching condition
\be
\Lambda_{A_r}^{2(r+1)}=(1+r)^2b^2 \Lambda_{A_{r-1}}^{2r},
\ee
after a rescaling and a shift of $y$ and $x$.
Since $\Lambda_{d,A_1}=\Lambda_{A_1}$
in the Pauli-Villas regularization scheme \cite{FINPOU,itsa},
we obtain the relation between the scales as \cite{itsa}
\be
\Lambda_{d,A_r}=2^{r-1}\Lambda_{A_r}.
\ee

The same procedure works for cases $B_r\rightarrow B_{r-1}$, 
$C_r\rightarrow C_{r-1}$ and $D_r\rightarrow D_{r-1}$,
if one considers the scalar vacuum expectation value (see section 2.3)
\be  
a=be_{1}+\hat{a}, \quad (\hat{a},e_{1})=0.
\ee
But a little care must be taken for the case $C_r \rightarrow C_{r-1}$.
This is because
there are two kinds of massive gauge bosons with different masses.
They are related to the roots $\sqrt2 e_1$ and 
$\frac1{\sqrt2}(e_i\pm e_1)\ (i=2,\cdots,r)$, respectively. 
Only the latter appears 
in the one-loop diagram of the renormalization of the gauge coupling of 
$C_{r-1}$. Hence the mass to be taken is $b$. 
We will give a list of the matching condition in the following.
\begin{itemize}
\item $B_r\rightarrow B_{r-1}$
\bea
\Lambda_{d,B_r}^{4r-2}&=&4 b^4 \Lambda_{d,B_{r-1}}^{4r-6},\CR
\Lambda_{B_r}^{4r-2}&=&b^4 \Lambda_{B_{r-1}}^{4r-6}.
\eea
\item $C_r\rightarrow C_{r-1}$
\bea
\Lambda_{d,C_r}^{2r+2}&=& b^2 \Lambda_{d,C_{r-1}}^{2r},\CR
\Lambda_{C_r}^{2r+2}&=& b^2 \Lambda_{C_{r-1}}^{2r}.
\eea
\item $D_r\rightarrow D_{r-1}$
\bea
\Lambda_{d,D_r}^{4r-4}&=& 4b^4 \Lambda_{d,D_{r-1}}^{4r-8},\CR
\Lambda_{D_r}^{4r-4}&=& b^4 \Lambda_{D_{r-1}}^{4r-8}.
\eea
\end{itemize}
All of the above relations for the dynamical scales $\Lambda_d$ are 
consistent with the microscopic instanton calculations (\ref{fonbr}),
(\ref{foncr}), (\ref{fondr}) and (\ref{finone}).

Now let us discuss the 
Higgs breakings $B_r\rightarrow A_{r-1}$ and $D_r\rightarrow A_{r-1}$.
Both of them can be realized by
\be  
a=b\sum_{i=1}^{r}e_{i}+\hat{a}, \quad
(\hat{a}, \sum_{i=1}^{r}e_{i})=0.
\ee
Similar discussions as above show that
\begin{itemize}
\item $B_r\rightarrow A_{r-1}$
\bea
\Lambda_{d,B_r}^{4r-2}&=&2^{3r-5}b^{2r-2}\Lambda_{d,A_{r-1}}^{2r}, \CR
\Lambda_{B_r}^{4r-2}&=&4^rb^{2r-2}\Lambda_{A_{r-1}}^{2r},
\eea
and
\item $D_r\rightarrow A_{r-1}$
\bea
\Lambda_{d,D_r}^{4r-4}&=&2^{3r-6}b^{2r-4}\Lambda_{d,A_{r-1}}^{2r},\CR
\Lambda_{D_r}^{4r-4}&=&2^{2r}b^{2r-4}\Lambda_{A_{r-1}}^{2r}.
\eea
\end{itemize}
Concerning $C_r$, we have the relation (\ref{btwandctw}) as well as 
$\Lambda_{d,C_2}=\Lambda_{d,B_2}$,
which should be trivially satisfied in the microscopic theory.
The relations in this subsection consistently fix the relations 
between $\Lambda_d$ and $\Lambda$ as follows: 
\bea
\Lambda_{d,B_r}^{4r-2}&=&2^{2r-7}\Lambda_{B_r}^{4r-2},\CR
\Lambda_{d,C_r}^{2r+2}&=&-2\Lambda_{C_r}^{2r+2},\CR
\Lambda_{d,D_r}^{4r-4}&=&2^{2r-8}\Lambda_{D_r}^{4r-4}.
\label{reldynsca}
\eea
Substituting (\ref{reldynsca}) into the results of the microscopic
one-instanton calculations (\ref{fonbr}), (\ref{foncr}) and (\ref{finone}),
we find they agree with the exact solutions (\ref{exabtw}),
(\ref{exabth}) for $B_{2,3}$ and (\ref{exac}) for $C_{2,3}$
on the numerical factors as well as on the functional forms.

\section{Conclusions and discussions}
In this paper, we have calculated the one-instanton correction to 
the prepotential in the $N=2$ SYM theory with any simple gauge group
by the microscopic instanton calculation,
and have found the microscopic one-instanton corrections agree with
those from the exact solutions for some classical gauge groups.
To derive the instanton corrections from the exact solutions, we have derived
the Picard-Fuchs equations from the proposed hyperelliptic curves and 
meromorphic differentials, and 
solve them to the one-instanton level near semi-classical regime.

The generalization to $N=2$ SQCD with fundamental hypermultiplets
would be straightforward. 
In $N=2$ $SU(2)$ SQCD with fundamental hypermultiplets,
the lowest instanton correction is the two-instanton effect \cite{sewi2}.
This is due to the $Z_2$ symmetry originating from the 
fact that the representation of $SU(2)$ is psudoreal.
For the other gauge groups, one naturally expects 
the lowest instanton correction will be the one-instanton contribution.
Hence it would be interesting to investigate the comparison in $N=2$
SQCD with the other gauge groups, especially for the similar cases
that some discrepancies have been reported \cite{AOYHAR,DORKHOQCDTWO}.
These cases would be generalized to 
 $N=2$ $SU(N_c)$ SQCD with $N_f$ ($2 N_c-2 \leq N_f \leq 2 N_c$)
fundamental hypermultiplets. 
In these cases, the result of the group integration 
would have a regular term, 
which can not be determined solely by the estimation of 
the poles. Thus one will have to improve the method in this direction. 

The investigation of the exceptional Lie groups would be also interesting.
We have already calculated the one-instanton correction to the
prepotential by the microscopic one-instanton calculation. Hence only the 
comparison with the exact solutions is necessary.
The explicit forms of the hyperelliptic curves for the $G_2$ \cite{alar},
$F_4$ and $E_{6,7,8}$ \cite{ABOALI,MaWa,EgHo} SYM theories 
have been already proposed. The new  feature of the $F_4$ and
$E_{6,7,8}$ cases is that
there are too many homology cycles on the hyperelliptic curves
compared to the number of the rank of the gauge groups.
The comparison would be an interesting non-trivial test for the validity
of the proposed curves.

\vskip5mm
\section*{Acknowledgments}

N.S. would like to thank K.~Higashijima, H.~Kanno, H.~Kunitomo,
T.~Nakatsu and M.~Sato for valuable discussions.
N.S. is supported by JSPS Research Fellowship for Young Scientists
(No.~06-3758).
The work of K.I. is supported in part by 
the Grant-in-Aid for Scientific Research from
the Ministry of Education (No.~08740188 and No.~08211209).

\appendix
\section*{Appendix A \ Conventions on the Cartan-Weyl basis}

We use the following conventions on the Cartan-Weyl basis
of the Lie algebra $\lg$ of a simple Lie group $G$.
\begin{eqnarray*}
[H_i,E_{\alpha}] &=& \alpha_i E_{\alpha}, \nonumber \\
{}[E_{\alpha},E_{-\alpha}] &=& \sum_{i=1}^r
\alpha_i H_i, \nonumber \\
\mbox{Tr}_{adj}(H_i H_j)&=&k_D \delta_{ij}, \nonumber \\
\mbox{Tr}_{adj}(E_\alpha E_{-\alpha'})&=&k_D \delta_{\alpha \alpha'},
\end{eqnarray*}
where $r$ is the rank of the Lie algebra $\lg$,
$\alpha_i$ is a root vector, and $k_D$ denotes the Dynkin index of 
the adjoint representation. The trace is taken in the adjoint representation.
The longest roots $\alpha^l$ are normalized by 
$
(\alpha^l,\alpha^l)=2.
$

\section*{Appendix B \ The Dynkin indices of the adjoint representation}
For completeness, we give here a list of Dynkin indices of the adjoint
representations of simple Lie algebras \cite{SLA}.
$$
\begin{array}{llll}
{k_D}^{A_r}=2(r+1), &  {k_D}^{B_r}=2(2r-1), & {k_D}^{C_r}=2(r+1), \\
{k_D}^{D_r}=4(r-1), & {k_D}^{G_2}=8, &  {k_D}^{F_4}=18, \\
{k_D}^{E_6}=24,&  {k_D}^{E_7}=36,& {k_D}^{E_8}=60. \\
\end{array}
$$

\end{document}